\documentclass{article}
\usepackage{ijcai21}

\usepackage{times}
\usepackage{soul}
\usepackage{url}
\usepackage[hidelinks]{hyperref}
\usepackage[utf8]{inputenc}
\usepackage[small]{caption}
\usepackage{graphicx}
\usepackage{amsmath}
\usepackage{amsfonts}
\usepackage{amsthm}
\usepackage{booktabs}
\usepackage{algorithm}
\usepackage{algorithmic}
\title{Exploring Fairness in District-based Multi-party Elections under different Voting Rules using Stochastic Simulations}

\author{
Adway Mitra
\affiliations
Indian Institute of Technology Kharagpur
\emails
adway.cse@gmail.com
}

\begin{document}

\maketitle

\begin{abstract}
Many democratic societies use district-based elections, where the region under consideration is geographically divided into districts and a representative is chosen for each district based on the preferences of the electors who reside there. These representatives belong to political parties, and the executive powers are acquired by that party which has a majority of the elected district representatives. In most systems, each elector can express preference for one candidate, though they may have a complete or partial ranking of the candidates/parties. We show that this can lead to situations where many electors are dissatisfied with the election results, which is not desirable in a democracy. The results may be biased towards the supporters of a particular party, and against others. Inspired by current literature on fairness of Machine Learning algorithms, we define measures of fairness to quantify the satisfaction of electors, irrespective of their political choices. We also consider alternative election policies using concepts of voting rules and rank aggregation, to enable voters to express their detailed preferences without making the electoral process cumbersome or opaque. We then evaluate these policies using the aforementioned fairness measures with the help of Monte Carlo simulations. Such simulations are obtained using a proposed stochastic model for election simulation, that takes into account community identities of electors and its role in influencing their residence and political preferences. We show that this model can simulate actual multi-party elections in India. Through extensive simulations, we find that allowing voters to provide 2 preferences reduces the disparity between supporters of different parties in terms of the election result.
\end{abstract}

\section{Introduction}
In a democracy, it is impossible to satisfy the preferences of all the people, all the time. Especially in multi-layered, multi-ethnic societies, the preferences of different people are often contradictory, and the democratic system often relies on numbers to decide which preferences to accept. However, this approach has a number of problems: various minority groups may remain permanently unrepresented or left out of the power structure, while political parties or other powerful players may game the system to maximize their benefits, even if it means that the preferences of a large number of electors are ignored. Such gaming may be through explicit steps like re-drawing the district boundaries (Gerrymandering) or through implicit steps like forging or breaking coalitions between different parties, setting up dummy candidates to wean away votes from opponent parties, and so on. To make the democratic process truly robust and egalitarian where everyone's preferences are respected as far as possible, it is necessary to explore alternate ways of eliciting people's preferences and aggregating them to choose representatives, without making the electoral process too cumbersome and opaque. It should be the aim of any democracy to give every person a direct or at least indirect access to the power structure and find a way for their voice to reach the decision makers, regardless of whether their opinions are finally accepted. This requires us to examine various electoral systems over a wide variety of social settings.

The aim of this work is to lay a framework to explore and evaluate various electoral settings, using concepts of Rank Aggregation and fairness in Machine Learning. We consider a number of voting rules and seat assignment policies, and define criteria to evaluate the fairness of these rules and policies to the electors. We also propose a stochastic model to simulate the results of elections under the different rules and policies discussed above, and evaluate them using the aforementioned fairness criteria. We show that the proposed model is sophisticated enough to capture various features of actual elections in multi-party electoral democracies like India. Finally, we use this model to generate a wide number of election scenarios (electors' preferences) and evaluate electoral policies over them. We point out the relative strengths and weaknesses of different policies.

\section{Related Work}
There is a huge body of work related to elections and voting rules in the context of Computational Social Choice Theory. Some works have pointed out how social cleavages influence the choice of residential areas \cite{d,e,ee} in many multi-cultural societies of the world. Mathematical models for such segregation have been proposed by~\cite{schelling1971dynamic},~\cite{grauwin2012dynamic}, where the residential preferences of agents in a social system are captured through stochastic processes based on utility functions. The political preferences and electoral choices are also often influenced by such social cleavages in democratic countries~\cite{b}. Mathematical models on how social influences can modulate the voting behaviors of individuals have been developed in~\cite{c},~\cite{jansen2013class}. 

District-based electoral systems often produced results that are not compatible with the popular support for different parties. In certain situations, a party with a higher popular support may win fewer seats that a less popular party. This is known as \emph{Referendum Paradox}, which occurs due to differences in geographical distributions of the supporters of the parties. This may happen naturally (as discussed in~\cite{f,2}) or due to deliberate tampering of district boundaries, popularly called Gerrymandering~\cite{1,3}. Various studies have tried to quantify the representation bias of elections~\cite{g,7}, and many mathematical models and algorithms have been proposed to alter the electoral system to make the results more competitive between parties and more representative of their popular support~\cite{g,2,3}, usually by re-defining districts~\cite{4,6} or by considering mobility of electors~\cite{5}. Recently, there was a study~\cite{mitra2020electoral} to develop stochastic models to simulate outcomes of district-based elections, taking into account the aforementioned factors like social cleavage and geographical distributions.

Separately, the question of ranking and rank aggregation, along with notions of fairness and bias have also been studied thoroughly in Machine Learning over the past decade. This problem usually deals with the situation where each user has a ranked preference list over a set of items, and an overall ranking has to be produced which is  unacceptable to the least number of users. A survey of initial rank aggregation methods is found in~\cite{lin2010rank}. A major aspect of this problem is how to compare different sets of complete and incomplete rankings~\cite{jagabathula2008inferring,pairwise,negahban2012iterative} or to evaluate any rank aggregation method by using criteria such as the Condorcet Criteria, or their statistical properties~\cite{rajkumar2014statistical}. Several rank aggregation algorithms are based on probabilistic models of permutations, such as the Placket-Luce and Mallows models and their extensions~\cite{lu2011learning},~\cite{guiver2009bayesian},~\cite{qin2010new},~\cite{volkovs2012flexible}. These ideas of rank aggregation have also been used in the context of computational social choice for comparing different voting and scoring rules for problems like predicting election winners from small samples~\cite{dey2015sample}, candidate unavailability~\cite{boutilier2014robust}, victory margin estimation~\cite{xia2012computing} and choosing multiple winners by proportional representation~\cite{lu2013multi}. 

A recent trend in Machine Learning is to quantify the fairness of different algorithms~\cite{feldman2015certifying}, in terms of biases in their predictions that may be explicit or implicit. The question of fairness in ranking and rank aggregation algorithms has been investigated recently~\cite{kuhlman2020rank,kuhlman2021measuring}, where the basic aim is to come up with an optimal ranking that will be sufficiently representative of different groups of candidates as identified by a protected attribute (eg. gender, race, nationality) ~\cite{zehlike2017fa,yang2017measuring2}. Statistical parity is considered as a desirable property to be maximized~\cite{yang2017measuring,asudeh2019designing}. The idea of fairness has also been used to promote representation and diversity in electoral outcomes~\cite{relia2021dire}.

\section{Motivating Examples}
First of all, let us illustrate a few situations, when district-based elections can throw up unfair results. Consider a population of 25, divided into 5 districts. There are two parties (A and B). Each person casts their vote for any one of the two parties, and the district representative is chosen by the plurality voting rule, i.e. the party which received the higher number of votes wins the corresponding seat. The number of votes obtained by the parties in the 5 districts in three scenarios is shown in Table~\ref{tab:eg1}. 

\begin{table}[]
    \centering
    \begin{tabular}{|c|c|c|c|c|c|}
    \hline
        - & D1 & D2 & D3 & D4 & D5 \\
    \hline
        A & \textbf{3}  & \textbf{3}  & \textbf{3}  & \textbf{3}  & \textbf{3} \\
        B & 2  & 2  & 2  & 2  & 2 \\
    \hline
    \end{tabular}
    \begin{tabular}{|c|c|c|c|c|c|}
    \hline
        - & D1 & D2 & D3 & D4 & D5 \\
    \hline
        A & \textbf{5}  & \textbf{4}  & \textbf{3}  & \emph{2}  & \emph{1} \\
        B & 0  & \emph{1}  & \emph{2}  & \textbf{3}  & \textbf{4} \\
    \hline
    \end{tabular}
    \begin{tabular}{|c|c|c|c|c|c|}
    \hline
        - & D1 & D2 & D3 & D4 & D5 \\
    \hline
        A & \textbf{5}  & \textbf{5}  & \emph{2}  & \emph{2}  & \emph{1} \\
        B & 0  & 0  & \textbf{3}  & \textbf{3}  & \textbf{4} \\
    \hline
    \end{tabular}
    \caption{2-party election over 5 districts D1,\dots,D5. In all cases, party A has 15 votes and party B has 10. But in the upper case Party A wins all 5 seats, in the middle case they win 3 seats, while in the lower case Party B wins 3 of the 5 seats. The voters whose choices are represented are shown in bold, those whose choices are indirectly represented through their favoured party are shown in italics}
    \label{tab:eg1}
\end{table}

In the first case, each district uniformly has 3 supporters of party A and 2 supporters of party B. Under plurality rule, all 5 seats are won by party A. The 10 electors who prefer party B are unrepresented and shut out of the system. They are completely powerless to influence the governing body. 

In the second case, the first 3 districts are represented by Party A and the other 2 by Party B. Here the 3 voters from D4 and D5 who preferred Party A, and the 3 voters from D2 and D3 who preferred Party B are unrepresented in their districts, but they indirectly have representation in the governing body as their favoured parties do have representatives from other districts. 

In the third case, the first two districts are represented by party A and the other 3 by party B. So here we find that the 5 voters from D3, D4, D5 who preferred party A are unrepresented in their district, but indirectly they do have representation in the governing body through the representatives of D1, D2 who are from party A. However, this is a curious situation where the governing body is dominated by the representatives of the party that has less supporters overall. 

Based on the above analysis, we find that the second scenario is most fair, while the first one is most unfair. In the second and third cases, no one is left out of the system. In the second case 6 people are indirectly represented, and they are from both parties in equal number, while in the third case only 5 people are indirectly represented, but all of them are from party A. So we can say that the first scenario is strongly unfair to supporters of party B, and the third is weakly unfair to supporters of party A. However all of these scenarios can arise with occurrence probabilities depending on the dynamics of the political process in this hypothetical society. If scenarios 1 or 3 do arise, the setup (plurality voting rule) is unable to prevent them. 

\begin{table}[]
    \centering
    \begin{tabular}{|c|c|c|c|c|c|c|}
    \hline
        - & V1 & V2 & V3 & V4 & V5 & V6\\
    \hline
       P1 & A  & A  & A  & B  & B  &  C\\
       P2 & B  & B  & C  & C  & C  &  B\\
       P3 & C  & C  & B  & A  & A  &  A\\
    \hline
    \end{tabular}
    \caption{The ranked list of 5 voters V1-V6 over 3 candidates A,B,C.}\label{tab:eg2}
\end{table}

In another example, consider 6 electors and 3 candidates. In this case, each candidate has a ranked list of preferences over the candidates. These preferences are shown in Table~\ref{tab:eg2}. We see that candidate A is a divisive candidate (either top or bottom choice), but in plurality voting system, candidate A will be elected, which will dissatisfy voters V4, V5 and V6. In case of Borda count, B will be elected with score 7, dissatisfying only V3. So in this scenario, Borda count is \emph{fairer} than plurality voting, despite its logistical disadvantage as each voter must specify a full ranking.

\section{Problem Formulation}
Consider a geographical region that is divided into $S$ districts, and corresponding to each district there is a \emph{seat} in the governing body like parliament. There are also $R$ extra seats, which may be filled up as necessary. $N$ number of electors are present in the region, partitioned into the districts as $n_1,n_2,\dots,n_S$, where $\sum_{i=1}^Sn_i=N$. Also, there are $K$ political parties, and in each district there are $K$ candidates sponsored by them (we ignore independent candidates). In the rest of this paper, we will use the terms \emph{party} and \emph{candidate} interchangeably.  

In any district, suppose $\mathcal{C}$ represents the set of candidates. Each elector may have a ranked list of preference on these candidates. Denote by $\mathcal{R(\mathcal{C})}$, the set of possible rankings (total order) over the candidates. Then a \emph{aggregation /scoring rule} is a function $\bigcup_{n,m\in \mathcal{Z}^{+}, m\leq |C|}\mathcal{R(\mathcal{C})}^n \mapsto \mathcal{C}^m$, i.e. it collects the rankings over the candidates from $n$ electors and computes a set of $m$ candidates, who are called the \emph{elected}. Usually, $m=1$. The party sponsoring the elected candidate of district $s$ is denoted by $w(s)$, and the number of district seats won by candidates from party $k$ is denoted as $V(k)$. Clearly, $\sum_{k=1}^KV(k)=S$. 

\subsection{Fairness Criteria}
Based on the previously mentioned motivating examples, we define fairness criteria for elections. For this purpose, we first define the following:
\begin{enumerate}
    \item \emph{Represented}: An elector $i$ from district $s$ is said to be \emph{represented}, if the candidate of their top choice $j=\mathcal{R}_i(\mathcal{C}_s)(1)$  is among the elected candidates from $s$, i.e. $j \in w(s)$
    \item \emph{Indirectly represented}: An elector $i$ from district $s$ is said to be \emph{indirectly represented} if the candidate of their top choice $j$  is not among the elected candidates from $s$, but belongs to the same party as at least one elected candidate from another district
    \item \emph{Unrepresented} An elector $i$ is \emph{unrepresented} if no elected candidate from any district belongs to the same party as the top choice of $i$
    \item \emph{K-Dissatisfied} An elector $i$ from district $s$ is \emph{dissatisfied} if they are unrepresented, and the elected candidates from $s$ are not within their top $K$ preferences
\end{enumerate}

Using these definitions, we define fairness criteria as follows:
\begin{enumerate}
    \item \emph{Net K-Dissatisfied ($ND(k)$)}: The total number of $k$-dissatisfied electors, across all districts. This number should be minimized.
    \item \emph{Net Unrepresented ($NUR$)}: The total number of unrepresented electors, across all districts. This number should be minimized, so that the opinions of as many people as possible have some access to the governing body.
    \item \emph{Partywise Indirect Representation Parity ($PIRP$)}: Let $IR(p)$ be the number of indirectly represented electors, whose candidates of first preference belong to party $p$. We define $PIRP$ score as the variance of $IR$ over all the parties, and this score should be minimized, so that the election results are not biased towards or against the supporters of any party.
    \item \emph{Net Represented ($NR$)}: The total number of \emph{represented} electors. This number should be maximized, as it indicates the net satisfaction and empowerment. 
    \item \emph{Partywise Representation Parity ($PRP$)}: Let $R(p)$ be the number of represented electors whose first preference is party $p$. We define $PRP$ score as the variance of $R$ over all the parties, and this score should be minimized, so that the election results are not biased towards or against the supporters of any party. 
    \item \emph{Net Borda Count ($NBC$)}: The Borda count of an elector $i$ for an elected candidate $j$ indicates the number of opponent candidates who ranks below $j$ in the ranked preference list of $i$. A high Borda count indicates the satisfaction of the elector. The Net Local Borda Count is the sum of the Borda counts of all electors for the candidates elected from their respective districts. This score should be maximized to indicate satisfaction of the electors over the elected candidates.
    \item \emph{Partywise Borda Score ($PBS$)}: Let $BC(p)$ indicate the Net Borda count of those electors whose first preference is party $p$. We define $PBS$ as the variance of $BC$ over all the parties. This score too be minimized, like the other partywise scores.

\end{enumerate}

\subsection{Voting and Aggregation Rules}
As already discussed, a number of voting rules have been proposed in social choice theory. However, these require the voters to specify their preferences over the given set of candidates as a ranked list. This is often difficult for electors, especially if there is a long list of candidates, which happens in multi-party democracies. So in such systems, electors are usually given the option of specifying only one choice. Here we consider a set of voting rules that may be used in actual elections, without making the process too cumbersome for electors. 

\begin{itemize}
    \item \emph{$k$-approval}: Each elector is allowed to specify $k$ choices. The standard rule is $1$-approval. For modeling purposes, it may be assumed that each voter has total or partial ranking over the candidates, of which they choose the top $k$. 
    \item \emph{Weighted $k$-approval}: Each elector is allowed to specify $k$ choices, but each choice has a weight, specified by a vector $\{\alpha_1,\alpha_2,\dots,\alpha_k\}$. 
    \item \emph{approval}: Each elector can either approve or disapprove each candidate. For modeling purposes, it may be assumed that the elector assigns a score to each candidate, and approves only those candidates whose score exceeds a cutoff.
    \item \emph{negative vote}: Each elector can cast a positive vote for, and/or a negative vote against one candidate. This is a restricted version of the \emph{approval} rule.
    \item \emph{transferable vote}: Each elector can cast one vote and also mark another candidate as second choice, to whom the vote will be transferred, in case the preferred candidate is unable to be elected.
\end{itemize}

Once the votes are collected from all the electors, it is necessary to aggregate them using \emph{scoring rules}. This may be done using various rank aggregation techniques, as already discussed. However, most of these techniques require each elector to provide a complete ranking over the candidates, which is not possible according to the voting rules mentioned above. So we adapt the scoring rules to suit the voting rules, as follows:

\begin{itemize}
    \item \emph{Plurality}: The positive or approval votes in favour of each candidate from all the electors are added up and sorted in descending order. This is possible for the $k-approval$ and $approval$ voting rules. The winners can be elected in the following ways: 
    \begin{itemize}
        \item The candidate with maximum number of votes is elected
        \item The top-$k$ candidates from the sorted list are elected, for a fixed $k$
        \item All candidates having a minimum number of votes are elected
    \end{itemize}
    \item \emph{Weighted Plurality}: In case of the weighted $k$-approval, all votes in favour of each candidate are added up according to their weights. The candidates are then sorted in the descending order of this weighted sum, and elected according to any of the 3 policies mentioned above
    \item \emph{Net Plurality}: This is applicable in case approval and negative vote rules. Each positive vote or approval is considered to have a weight of $+1$, while each negative vote has a weight of $-1$. Weighted plurality scoring rule (mentioned above) is applied using these weights.
    \item \emph{Plurality with Transfer}: This scoring rule is applied for transferable vote. The votes for each candidate are first counted, and the candidates ranked in descending order of votes polled. Then the candidate with least number of votes is eliminated, and the votes in their favor are transferred as specified by the corresponding electors. The remaining candidates are re-ordered according to the direct and transferred votes. The process continues till candidates can be elected according to any of the aforementioned ways.
\end{itemize}

Each (voting rule, scoring rule) combination will be referred to as \emph{electoral policy}. Having defined these voting and scoring rules, we wish to evaluate them according to the fairness criteria defined above. It is likely that in different scenarios regarding the choices of the electors, the most fair rule will be different. However, not all scenarios are equally likely. To make an average-case analysis, we use a Monte Carlo approach where scenarios will be sampled from an election simulation model, the results computed according to various voting and scoring rules, and their fairness scores measured accordingly.

\section{Election Simulation Model}
First of all, we need a model to simulate the results of an election in a multi-party, district-based setting. It has already been discussed how social identities and connections influence the districts of residence, political preference and final voting decision of individual electors. As a result, the result of district-based election, which is understood in terms of the number of seats won by the different parties, is sensitive to a number of factors beyond the overall popularity of the different parties or candidates. Below, we discuss a stochastic model for simulating such elections.

\subsection{Model Definition}

We have $K$ parties competing over $S$ seats corresponding to districts, in which reside $N$ people. Assume that there are $C$ social communities, and $\theta_c$ denotes the proportion of people from community $c$. $\theta$ is sampled from a Stick-breaking prior. To every person $i$, we assign their community as $C(i) \sim Categorical(\theta)$. The people from the same community tend to stay together in the same district. Each person $i$ is assigned to district $S(i)$ by following a Chinese Restaurant Process~\cite{crp} with parameter $\alpha$, where each district is considered to be a table. Person $i$, resides in district $s$ with probability proportional to $\alpha n_s(C(i)) = \alpha\sum_{j=1}^{i-1}\mathbb{I}(C(j)=n)\mathbb{I}(S(j)=s)$ (i.e. number of people from same community as $i$ already residing in district $s$), or resides in any district chosen uniformly at random with probability proportional to $(1-\alpha)$. However, through book-keeping we constrain the above process so that all districts have equal population. This ensures that for each community, certain districts turn into strongholds. Figure 1 shows the histograms of the district-wise distribution of a community generated in this way. 

\begin{figure}
    \centering
    \includegraphics[width=1.8in,height=1.5in]{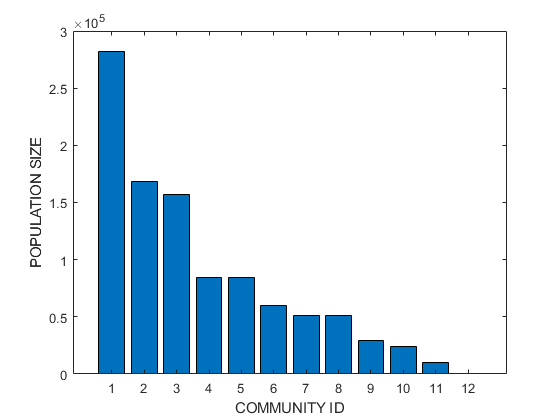}\includegraphics[width=1.8in,height=1.5in]{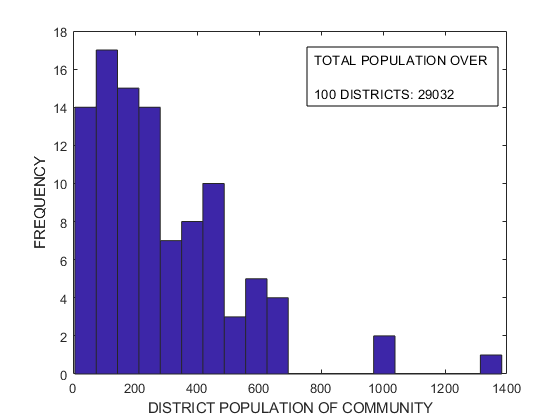}
    \caption{Top: sizes of 12 communities created by Stick-Breaking Process (Total population=1 Million), Bottom: District-wise distribution of one community's population}
    \label{fig:pop_distr}
\end{figure}

Each community is associated with a prior over the political preferences of its members. For community $c$ and party $k$, we assign $\phi_{ck} \in \{-1,0,1\}$, indicating if the relation between them is bad (-1), neutral (0) or good (1). The values $\phi_{ck}$ are sampled uniformly at random, with the constraint that no party can have good relation with more than half of the total electorate. Also, a variance $\sigma_k$ is associated with each party which may be drawn from a Gamma distribution. Finally, for each elector $i$, their valuation of party $k$ is denoted by $\lambda_{ik} \sim \mathcal{N}(\phi_{ck},\sigma_k)$ where $c=C(i)$. A party with high $\sigma$ is strongly liked by some and strongly disliked by others (indicating its ``polarizing" nature). Clearly, this valuation $\lambda_{ik}$ can be either positive or negative.

Next, in an election each elector casts their votes on the basis of these valuations $\lambda$, according to the voting rules. In case of $k$-\emph{approval} and \emph{weighted $k$-approval}, each voter chooses the parties according to their top $k$ valuations. In case of \emph{approval}, each elector approves the candidates from the parties with positive valuations, and disapproves the rest. In case of \emph{negative vote}, the party with the least valuation gets the negative vote.

In actual elections, electors rarely vote according to their individual inclinations. They are also influenced their social network. We consider another version of the model (Local Influence), where the $i$-th elector combines their own valuations $\lambda_{ik}$ with the mean valuations of other electors in the same district, as $\hat{\lambda}_{ik}=\kappa\lambda_{ik}+(1-\kappa)\bar{\lambda}_{ik}$ where $\bar{\lambda}_{ik}=\frac{\sum_{j=1}^N\mathbb{1}(S(j)=S(i))\lambda_{jk}}{\sum_{j=1}^N\mathbb{1}(S(j)=S(i))}$, and $\kappa \sim Beta(a,b)$. Influences on an elector need not be local only, it is possible to consider overall and community-wise influence, or the social network of each specific elector.

In a nutshell, the election model may be written as:
\begin{eqnarray}\label{eq:1}
&\theta \sim SBP(c) \nonumber \\
&\phi_{ck} \sim Uniform\{-1,0,1\} \forall c,k \nonumber \\ 
&C(i) \sim Categorical(\theta) \forall i\in\{1,N\} \nonumber \\
&S(i) \sim CRP(C(i),\alpha) \forall i\in\{1,N\} \nonumber \\
&\sigma_k \sim Gamma(c) \forall k \nonumber \\
&\lambda_{ik} \sim \mathcal{N}(\phi_{ck},\sigma_k) \text{ where } c=C(i), \forall i,k \nonumber \\
&\kappa \sim Beta(a,b), \hat{\lambda}_{ik}=\kappa\lambda_{ik}+(1-\kappa)\bar{\lambda}_{ik} \forall i,k \nonumber \\ 
&\text{ where } \bar{\lambda}_{ik}=\frac{\sum_{j=1}^N\mathbb{1}(S(j)=S(i))\lambda_{jk}}{\sum_{j=1}^N\mathbb{1}(S(j)=S(i))}  
\end{eqnarray}

\subsection{Model Evaluation}
It is important to validate the above model to show that it is capable of producing realistic results. For this purpose, we attempt to simulate actual elections in India - a multi-party democracy. The  election results in India are available at \url{https://eci.gov.in/statistical-report/statistical-reports/}.

We carry out two experiments under different settings. In the first experiment, we consider Delhi National Capital Region- a small state assembly with 70 seats. Roughly 9 million people participate in the elections that are primarily between 3 major political parties. We consider $C=5$ arbitrary communities, and the party-community relations are generated randomly as mentioned above. A large number of election scenarios $\{S(i),\{\lambda_{ik}\}_{k=1}^K\}_{i=1}^N$ are simulated from the model, from which we try to retrieve the results that are closest to each of the past 5 elections in the region. The simulated and actual election results are compared based on proportions of popular votes and number of seats won by the 3 parties. The results are shown below in Table 3. We find that the model can produce simulations that are reasonably close to the actual results. It must be remembered that in a district-based election, the mapping from popular vote distribution to seat distribution is many-to-many.

\begin{table}[]
    \centering
    \begin{tabular}{|c||c|c|c||c|c|c|}
    \hline
    \multicolumn{1}{|c||}{} & \multicolumn{3}{|c||}{observed} &  \multicolumn{3}{|c|}{simulated}\\
    \hline
    Year & M1 & M2 & M3 & M1 & M2 & M3\\
    \hline 
    2013  & 0.37 & 0.34 & 0.29 & 0.36 & 0.35 & 0.29 \\
    2014  & 0.48 & 0.35 & 0.17 & 0.44 & 0.39 & 0.17 \\
    2015  & 0.56 & 0.34 & 0.10 & 0.59 & 0.26 & 0.15 \\
    2019  & 0.58 & 0.23 & 0.19 & 0.59 & 0.26 & 0.15 \\
    2020  & 0.55 & 0.40 & 0.05 & 0.53 & 0.34 & 0.14 \\
    \hline
    Year & V1 & V2 & V3 & V1 & V2 & V3\\
    \hline 
    2013  & 32 & 30 & 8 & 33 & 29 & 8 \\
    2014  & 60 & 10 & 0 & 56 & 14 & 0 \\
    2015  & 67 &  3 & 0 & 62 & 8  & 0 \\
    2019  & 65 &  5 & 0 & 62 & 8 & 0 \\
    2020  & 62 &  8 & 0 & 58 & 12 & 0 \\
    \hline
    \end{tabular}
    \caption{Comparison of observed and simulated results for past 5 elections in Delhi-NCR. Above: rounded popular vote shares (M1,M2,M3) of 3 main parties, below: seats won (V1,V2,V3) by these parties. Here 1,2,3 are not specific parties but simply those placed first, second and third in terms of outcome of each election.}
    \label{tab:my_label}
\end{table}

In another experiment, we consider two elections held in another Indian state of Odisha, which has 147 seats. Roughly 23 million people participated in another tri-partite contest. In this case, we had an estimate of the preferences for the 3 parties in 5 social communities on the basis of post-poll surveys\footnote{\url{https://www.thehindu.com/elections/lok-sabha-2019/naveens-track-record-helps-to-overcome-bjp-blitz/article27267792.ece}}. The $\theta$ and $\phi$ matrices are accordingly specified before-hand. It turns out that the popular vote proportions and seat proportions, as simulated by the models, are reasonably close enough to the actual results, as shown in Table 4. This shows that our models can simulate realistic results. The results shown are in the individual-based version of the model, without considering local influence ($\lambda$). In presence of $\lambda$, we see that the seat proportion of different parties is closer to the popular vote proportion than the observations. 

\begin{table}[]
    \centering
    \begin{tabular}{|c||c|c|c||c|c|c|}
    \hline
    \multicolumn{1}{|c||}{} & \multicolumn{3}{|c||}{observed} &  \multicolumn{3}{|c|}{simulated}\\
    \hline
    Year & M1 & M2 & M3 & M1 & M2 & M3\\
    \hline 
    2019-1  & 0.48 & 0.34 & 0.18 & 0.51 & 0.31 & 0.18 \\
    2019-2  & 0.45 & 0.40 & 0.15 & 0.43 & 0.41 & 0.16 \\
    \hline
    Year & V1 & V2 & V3 & V1 & V2 & V3\\
    \hline 
    2019-1  & 114 & 23 & 10 & 113 & 28 & 6 \\
    2019-2  & 88  & 52 & 7  & 89 & 58 & 0 \\
    \hline
    \end{tabular}
    \caption{Comparison of observed and simulated results for 2 simultaneous elections in Odisha state, 2019. Above: rounded popular vote shares (M1,M2,M3) of 3 main parties, below: seats won (V1,V2,V3) by these parties. Here 1,2,3 are not specific parties but simply those placed first, second and third in terms of outcome of each election.}
    \label{tab:my_label}
\end{table}

\section{Simulation for Policy Evaluation}
Having defined the simulation model, we now proceed with the simulation under a variety of settings. For all the following simulations, we consider a population of $N=1 million$, spread over $S=100$ districts.

\subsection{A Motivating Example}

First of all, let us consider a motivating example where there are $C=3$ communities and $K=3$ parties. The distribution of the community sizes is $\theta=\{0.5,0.3,0.2\}$. Consider party 1 which is majoritarian, i.e. it patronizes the first (largest) community while victimizing the third (smallest) community, i.e. $\phi_1=\{1,0,-1\}$. The third party represents the interests of the smallest community but is hated by the largest community $\phi_3=\{-1,0,1\}$ while the centrist second party tries to balance everyone's interest $\phi_2=\{0,0,0\}$. We also consider that the first and third parties have higher variance $\sigma_1=\sigma_3=2$ than the second party $\sigma_2=1$.

We consider the results of such an election under the different voting and scoring rules as already discussed in Section 4.2. The number of seats won by the 3 parties and the different fairness scores as defined in Section 4.1, (average over 10 runs) are presented in Table 3. The table shows the results where each elector votes individually (without local influence), but the results are near-identical either way.

\begin{table}[]
    \centering
    \begin{tabular}{|c|c|c|c|c|c|c|}
    \hline
      &  1-app & 2-app & wtd & app & neg & trns\\
      &        &     & 2-app &    & vote & vote\\
      \hline
      v(1) & 70 & 38 & 69 & 69 & 69 & 69\\
      v(2) & 0 & 56 & 1 & 1 & 0 & 0\\
      v(3) & 30 & 6 & 30 & 30 & 30 & 30\\
      NR & 45 & 41 & 45 & 45 & 45 & 45\\
      NBC & 1.09 & 1.09 & 1.09 & 1.09 & 1.09 & 1.08\\
      ND(2) & 27 & 27 & 27 & 27 & 27 & 27\\
      PRP & 3.9 & 2.9 & 3.8 & 4.0 & 4.0 & 3.7 \\
      PBS & 0.4 & 0.32 & 0.37 & 0.4 & 0.4 & 0.36\\
      \hline
    \end{tabular}    
    \caption{Analysis of the synthetic setting election under different electoral policies.}
    \label{tab:my_label}
\end{table}

We find that under most policies, the majoritarian party wins the election with about 70 seats, while the subaltern party too punches above its weight, winning about 30 seats. The centrist party doesn't make any impact. However, in case of 2-approval policy, the centrist party wins the election. In that case, the percentage of the electorate directly represented is the least, though the variance among the supporters of the three parties in terms of representation and Borda score of satisfaction is also the minimum. Thus, double-choice voting with equal weightage to the second choice creates a more ``middle-of-the-ground" result which tries to accommodate the different communities.

\subsection{Randomized Community Settings}
Next, we consider simulations where the community-based variables $\theta$ and $\phi$ are randomized, as in Eq~\ref{eq:1}. We consider 4 cases: $(C=4,K=3)$, $(C=6,K=4)$, $(C=10,K=4)$ and $(C=12,K=4)$. In each case, we run simulations over 50 randomized settings of $\theta$ and $\phi$, with 5 runs in each setting (total 250 runs in each case, 1000 simulations overall). The district assignments of the electors and their valuations of the different parties are generated separately for each run. In each run, we compute election results according to the different voting and scoring rules as discussed in Section 4.2. As we saw, in most cases there is a one-to-one mapping between voting and scoring rules (eg. $k$-approval $\mapsto$ $k$-plurality). We consider $k=1$ and $k=2$ for $k-approval$, with $\{\alpha_1=1,\alpha_2=1\}$ and $\{\alpha_1=1,\alpha_2=0.5\}$ for $2$-approval. We finally calculate the mean statistics of the different fairness measures (Section 4.1) over all the settings for each electoral policy, and use them to compare the different policies.

\begin{itemize}
    \item \textbf{Net Represented}: We compare the percentage of the electorate who are directly represented under each electoral policy, for the 4 cases of $(C,K)$. The results for  are shown in Table 6 by considering the mean over a hundred simulations. The results show that in most settings, the standard form of voting ($1$-approval) is the best, though the differences between the different policies are not large. However, the mean values hide some variance (as shown in the box-plots of Fig. 2). We also calculate the number of simulation runs in which each policy returned the best results. We find that \textbf{$1$-approval} is  the best policy for both voting models, though under the local influence model it is closely followed by transferable vote and approval policies.

    \item \textbf{Net 2-Dissatisfied}  We compare the percentage of the electorate who are dissatisfied with the results, i.e. the winner from their district does not occur within their top 2 preferences. Again, this is studied under each electoral policy, for the 4 cases of $(C,K)$. The results are shown in Table 7, which show that on average (over hundred runs) $1-approval$ is the quite poor, though none of the alternative policies is consistently on top. By calculating the best performer in each run, we find that \textbf{$2$-approval}, i.e. double-choice voting with equal weightage is the best policy for both individual and local influence voting models. The variance in the performance of the policies over the runs are shown in the box-plot of Fig. 3. The least values are usually attained by $2$-approval.
    
    \item \textbf{Net Borda Count} Next, we compare the mean Borda score of each elector with respect to the final result, i.e. the mean position of each elected representative in their ranked list of preferences. High Borda score indicates high level of satisfaction with the result. The results are shown in Table 8, which show very little difference between the policies. On considering the best policies in each run, it turns out that $1$-approval and $2$-approval are at par. The performances of the policies over the runs are shown in the box-plot of Fig.4, which shows that the best results are achieved by 2-approval and negative voting. 
    
    \item \textbf{Partywise Representation Parity} Next, we compare the variance in the number of represented supporters of the different parties. The results are shown in Table 9, which show that in most cases the standard $1$-approval creates a high variance among the represented supporters of different parties, i.e. tends to over-represent one party's supporters at the cost of others. This is best mitigated by \textbf{$2$-approval}. These results are corroborated on considering the best performing policy of each run.
    
    \begin{table}[]
        \centering
        \begin{tabular}{|c|c|c|c|c|c|c|}
            \hline
              (C,K) &  1-app & 2-app & wtd & app & neg & trns\\
              & & & 2-app & & vote & vote\\
            \hline
              $4,3$ & \textbf{39.24} & 38.04 & 39.05 & 38.89 & 39.22 & 39.16\\
              $6,4$ & \textbf{29.20} & 28.44 & 29.10 & 28.35 & 28.83 & 28.98\\
              $10,4$ & \textbf{28.44} & 27.54 & 28.12 & 27.59 & 27.91 & 28.14\\
              $12,4$ & \textbf{28.82} & 28.42 & 28.77 & 28.44 & 28.63 & \textbf{28.82}\\
            \hline
              $4,3$ & \textbf{43.89} & 42.86 & 43.67 & 43.61 & 43.76 & 43.83\\
              $6,4$ & 32.41 & 31.92 & \textbf{32.53} & 31.80 & 32.40 & 32.29\\
              $10,4$ & \textbf{31.76} & 31.01 & 31.61 & 31.20 & 31.67 & 31.45\\
              $12,4$ & \textbf{32.69} & 32.38 & 32.44 & 32.43 & 32.51 & 32.65\\ 
            \hline
        \end{tabular}
        \caption{Mean Percentage of the electorate who are directly represented under different electoral policies. Above: Individual voting model, Below: Local influence model. (app: approval, wtd: weighted, neg: negative, trns: transferable)}
        \label{tab:netrep}
    \end{table}
    
    \begin{figure}
    \centering
    \includegraphics[width=1.8in,height=1.8in]{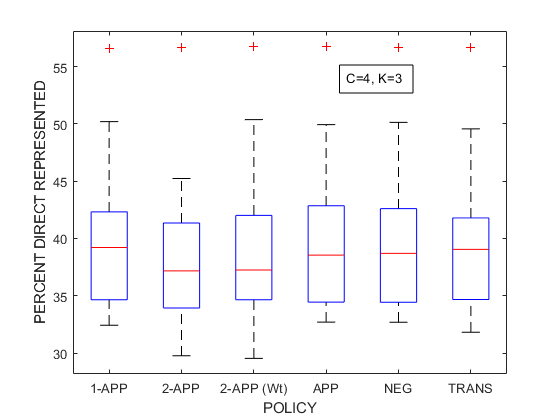}\includegraphics[width=1.8in,height=1.8in]{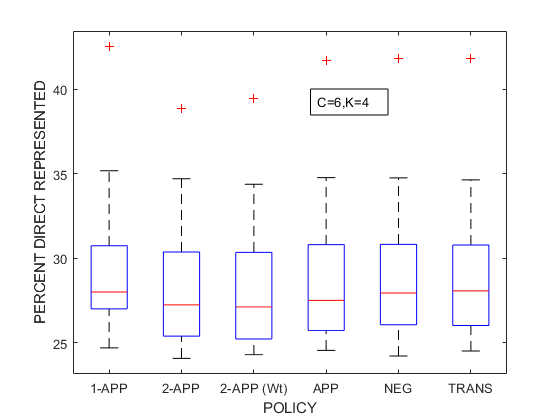}
    \includegraphics[width=1.8in,height=1.8in]{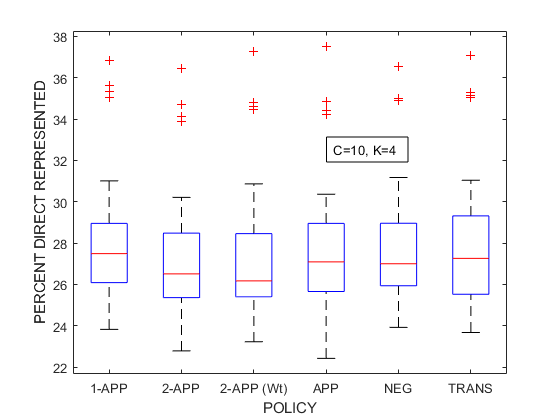}\includegraphics[width=1.8in,height=1.8in]{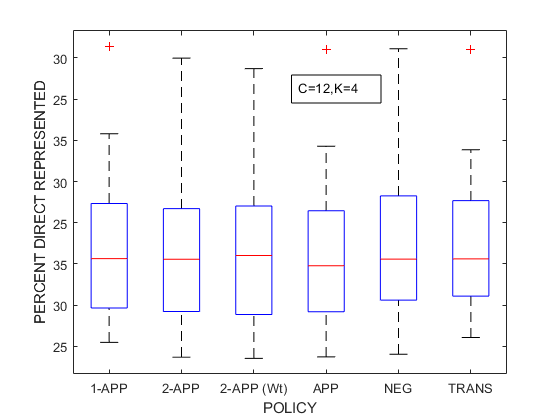}    
    \caption{Box-plots of ``Net Represented" measure for different policies, in different $(C,K)$ settings for individual voting model}
    \label{fig:pop_distr}
    \end{figure}
    
    \item \textbf{Partywise Borda Score} Next, we compare the variance in the Net Borda score of the supporters of the different parties. The results are shown in Table 10. Just like Partywise Representation Parity, it turns out that the standard $1$-approval creates disparity where the supporters of one party are more satisfied than the rest, while the situation is improved under most other policies, though most frequently under \textbf{$2$-approval}. However, $1$-approval works best in the presence of many communities.
    
    \begin{table}[]
        \centering
        \begin{tabular}{|c|c|c|c|c|c|c|}
            \hline
              (C,K) &  1-app & 2-app & wtd & app & neg & trns\\
              & & & 2-app & & vote & vote\\
            \hline
              $4,3$ & 29.63 & 29.00 & 28.79 & 28.94 & \textbf{28.77} & 28.90\\
              $6,4$ & 45.84 & \textbf{45.14} & 45.18 & 45.29 & 45.60 & 45.46\\
              $10,4$ & 46.41 & 46.03 & 46.11 & \textbf{46.02} & 46.25 & 46.10\\
              $12,4$ & 46.32 & \textbf{45.29} & 45.77 & 45.32 & 45.84 & 45.79\\
            \hline
              $4,3$ & 25.63 & 25.91 & 25.48 & 25.48 & 25.48 & 25.54\\
              $6,4$ & 41.87 & 41.39 & \textbf{41.34} & 41.57 & 41.67 & 41.73\\
              $10,4$ & 42.61 & 42.73 & 42.53 & \textbf{42.48} & 42.53 & 42.66\\
              $12,4$ & 41.57 & \textbf{41.03} & 41.50 & 40.99 & 41.56 & 41.40\\
            \hline
        \end{tabular}
        \caption{Mean Percentage of the electorate who are 2-dissatisfied under different electoral policies. Above: Individual voting model, Below: Local influence model. (app: approval, wtd: weighted, neg: negative, trns: transferable)}
        \label{tab:netdiss2}
    \end{table}
    
    \begin{figure}
    \centering
    \includegraphics[width=1.8in,height=1.8in]{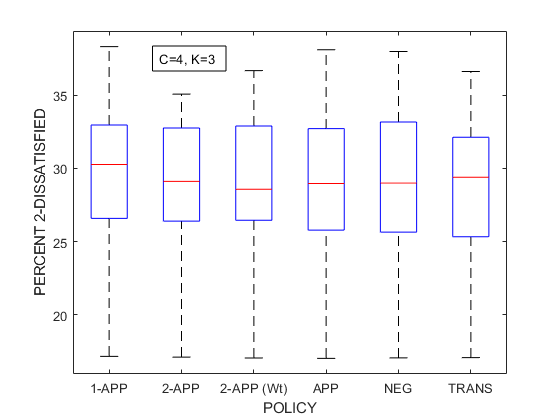}\includegraphics[width=1.8in,height=1.8in]{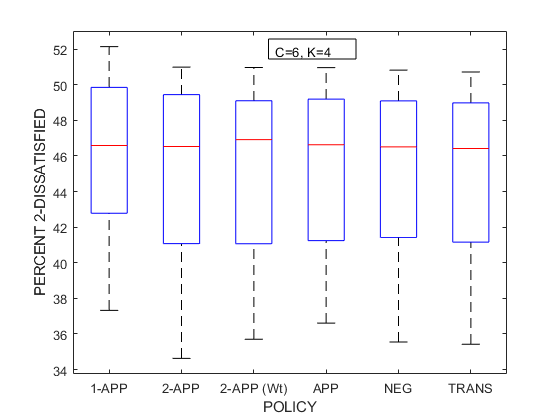}
    \includegraphics[width=1.8in,height=1.8in]{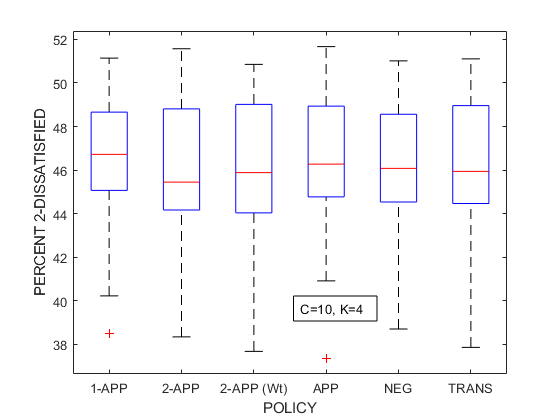}\includegraphics[width=1.8in,height=1.8in]{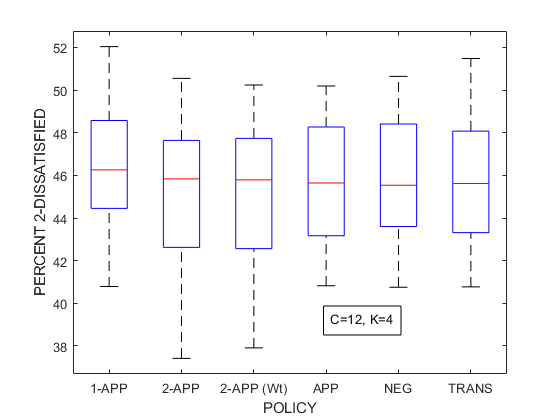}    
    \caption{Box-plots of ``2-dissatisfied" measure for different policies, in different $(C,K)$ settings for individual voting model}
    \label{fig:pop_distr}
    \end{figure}

    \begin{table}[]
        \centering
        \begin{tabular}{|c|c|c|c|c|c|c|}
            \hline
              (C,K) &  1-app & 2-app & wtd & app & neg & trns\\
              & & & 2-app & & vote & vote\\
            \hline
              $4,3$ & 1.05 & 1.05 & 1.05 & 1.05 & 1.05 & 1.05\\
              $6,4$ & 1.55 & 1.56 & 1.56 & 1.56 & 1.56 & 1.56\\
              $10,4$ & 1.55 & 1.55 & 1.55 & 1.55 & 1.55 & 1.55\\
              $12,4$ & 1.55 & 1.57 & 1.56 & 1.57 & 1.56 & 1.56\\
            \hline
              $4,3$ & 1.09 & 1.08 & 1.10 & 1.09 & 1.09 & 1.09\\
              $6,4$ & 1.6 & \textbf{1.62} & 1.61 & 1.61 & 1.61 & 1.61\\
              $10,4$ & 1.59 & 1.59 & 1.59 & 1.59 & 1.59 & 1.58\\
              $12,4$ & 1.62 & \textbf{1.63} & 1.62 & \textbf{1.63} & 1.62 & 1.62\\    
            \hline
        \end{tabular}
        \caption{Mean Net Borda Count, i.e. mean position of each elected representative in each elector's ranked preference list. Above: Individual voting model, Below: Local influence model. (app: approval, wtd: weighted, neg: negative, trns: transferable)}
        \label{tab:netdiss2}
    \end{table}
    
    \begin{table}[]
        \centering
        \begin{tabular}{|c|c|c|c|c|c|c|}
            \hline
              (C,K) &  1-app & 2-app & wtd & app & neg & trns\\
              & & & 2-app & & vote & vote\\
            \hline
              $4,3$ & 2.08 & \textbf{1.79} & 2.01 & 2.00 & 2.02 & 2.01\\
              $6,4$ & 0.59 & \textbf{0.54} & 0.57 & 0.56 & 0.56 & 0.56\\
              $10,4$ & 0.61 & \textbf{0.51} & 0.55 & 0.54 & 0.52 & 0.55\\
              $12,4$ & 0.68 & 0.70 & 0.68 & 0.68 & 0.69 & 0.68\\
            \hline
              $4,3$ & 3.04 & \textbf{2.8} & 3.03 & 2.99 & 3.1 & 3.06\\
              $6,4$ & 0.88 & 0.87 & 0.88 & \textbf{0.85} & 0.88 & 0.88\\
              $10,4$ & 0.85 & \textbf{0.79} & 0.83 & 0.84 & 0.82 & 0.82\\
              $12,4$ & 1.05 & 1.08 & 1.06 & \textbf{1.04} & 1.05 & 1.06\\    
            \hline
        \end{tabular}
        \caption{Mean Partywise Representation Parity, i.e. variance in the number of represented supporters of different parties. Above: Individual voting model, Below: Local influence model. (app: approval, wtd: weighted, neg: negative, trns: transferable)}
        \label{tab:netdiss2}
    \end{table}
    
    \begin{table}[]
        \centering
        \begin{tabular}{|c|c|c|c|c|c|c|}
            \hline
              (C,K) &  1-app & 2-app & wtd & app & neg & trns\\
              & & & 2-app & & vote & vote\\
            \hline
              $4,3$ & 0.25 & \textbf{0.22} & 0.23 & 0.24 & 0.24 & 0.23\\
              $6,4$ & 0.23 & \textbf{0.21} & 0.22 & 0.22 & 0.22 & 0.22\\
              $10,4$ & 0.22 & 0.21 & 0.21 & 0.22 & \textbf{0.20} & 0.21\\
              $12,4$ & 0.24 & 0.26 & 0.24 & 0.26 & 0.25 & 0.24\\
            \hline
              $4,3$ & 0.24 & \textbf{0.22} & 0.23 & 0.23 & 0.24 & 0.23\\
              $6,4$ & 0.23 & 0.22 & 0.22 & 0.22 & 0.22 & 0.23\\
              $10,4$ & 0.21 & 0.21 & 0.20 & 0.21 & 0.20 & 0.21\\
              $12,4$ & \textbf{0.23} & 0.25 & 0.25 & 0.24 & 0.24 & 0.24\\            
            \hline
        \end{tabular}
        \caption{Mean Partywise Borda Score, i.e. variance of net Borda score of supporters of different parties. Above: Individual voting model, Below: Local influence model. (app: approval, wtd: weighted, neg: negative, trns: transferable)}
        \label{tab:netdiss2}
    \end{table}
\end{itemize}

We find that in general the standard $1$-approval approach gives direct representation to the largest number of electors, but also increases disparity between the supporters of different parties, and the largest number of people are also dissatisfied under this policy. The $2$-approval policy is generally the perfect complement of $1$-approval, while the other policies like approval, negative vote and transferable vote fail to perform consistently well with respect to any of the considered measures. However, it should be noted that the variance between results of the different simulation runs is high (as indicated by Figs 2,3,4) and the difference between mean performance of different policies is generally low.

    \begin{figure}
    \centering
    \includegraphics[width=1.8in,height=1.5in]{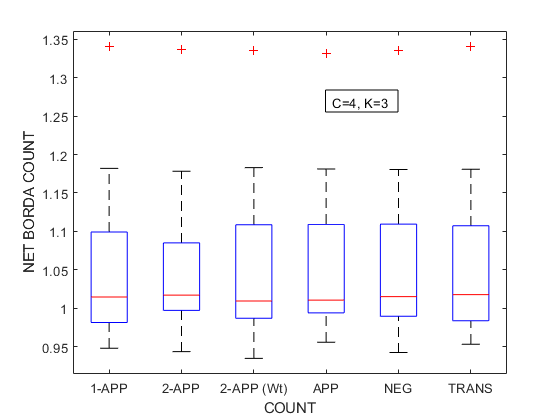}\includegraphics[width=1.8in,height=1.5in]{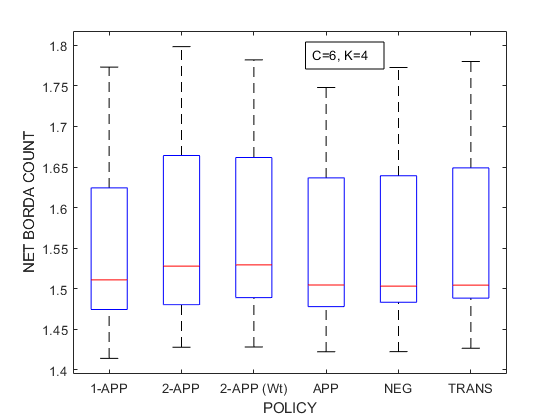}
    \includegraphics[width=1.8in,height=1.5in]{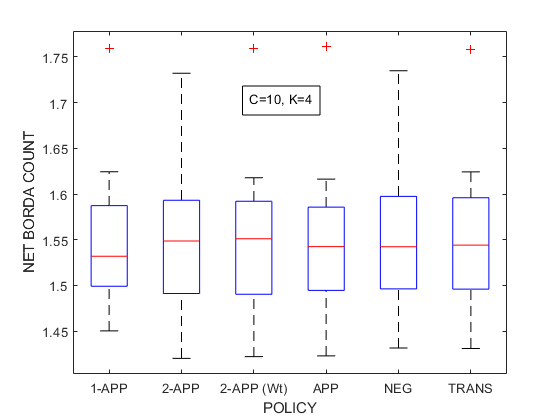}\includegraphics[width=1.8in,height=1.5in]{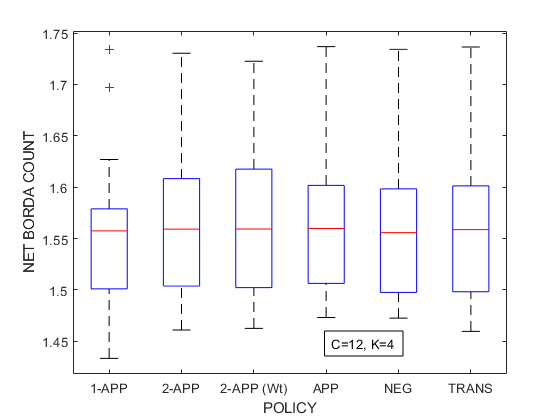}    
    \caption{Box-plots of ``Net Borda Count" measure for different policies, in different $(C,K)$ settings for individual voting model}
    \label{fig:pop_distr}
    \end{figure}

\section{Conclusion}
In this paper, we defined a theoretical framework to compare different electoral policies (voting and aggregation rules) over a number of measures that represent the satisfaction of electors from different political parties with the election outcome. These measures are inspired from fairness measures that are being used in recent Machine Learning literature. To explore the performance of the different policies, we proposed a stochastic model to simulate elections, where we bring in effects like community-based residence and political preference, and influence of neighbors on voting decision. We validated this model on actual election results in India. We found that the common $1$-approval rule which allows each elector to specify one choice, often creates disparities between the supporters of different parties, and the best way to prevent this seems to be allowing two choices with equal weightage. However, the intra-policy variance across simulation runs are large, and inter-policy differences are often small, suggesting the need to expand the simulation range so that a wider range of settings and scenarios can be explored, along with posterior distributions over the results.

\bibliographystyle{named}

\begin{thebibliography}{}

\bibitem[\protect\citeauthoryear{Asudeh \bgroup \em et al.\egroup
  }{2019}]{asudeh2019designing}
Abolfazl Asudeh, HV~Jagadish, Julia Stoyanovich, and Gautam Das.
\newblock Designing fair ranking schemes.
\newblock In {\em Proceedings of the 2019 International Conference on
  Management of Data}, pages 1259--1276, 2019.

\bibitem[\protect\citeauthoryear{Bachrach \bgroup \em et al.\egroup }{2016}]{7}
Yoram Bachrach, Omer Lev, Yoad Lewenberg, and Yair Zick.
\newblock Misrepresentation in district voting.
\newblock In {\em IJCAI}, pages 81--87, 2016.

\bibitem[\protect\citeauthoryear{Bharathi \bgroup \em et al.\egroup
  }{2018}]{ee}
Naveen Bharathi, Deepak~V Malghan, and Andaleeb Rahman.
\newblock Isolated by caste: Neighbourhood-scale residential segregation in
  indian metros.
\newblock {\em IIM Bangalore Research Paper}, (572), 2018.

\bibitem[\protect\citeauthoryear{Borodin \bgroup \em et al.\egroup }{2018}]{2}
Allan Borodin, Omer Lev, Nisarg Shah, and Tyrone Strangway.
\newblock Big city vs. the great outdoors: Voter distribution and how it
  affects gerrymandering.
\newblock In {\em IJCAI}, pages 98--104, 2018.

\bibitem[\protect\citeauthoryear{Boutilier \bgroup \em et al.\egroup
  }{2014}]{boutilier2014robust}
Craig Boutilier, J{\'e}r{\^o}me Lang, Joel Oren, and H{\'e}ctor Palacios.
\newblock Robust winners and winner determination policies under candidate
  uncertainty.
\newblock In {\em Proceedings of the AAAI Conference on Artificial
  Intelligence}, volume~28, 2014.

\bibitem[\protect\citeauthoryear{Braha and De~Aguiar}{2017}]{c}
Dan Braha and Marcus~AM De~Aguiar.
\newblock Voting contagion: Modeling and analysis of a century of us
  presidential elections.
\newblock {\em PloS one}, 12(5):e0177970, 2017.

\bibitem[\protect\citeauthoryear{Brooks \bgroup \em et al.\egroup }{2006}]{b}
Clem Brooks, Paul Nieuwbeerta, and Jeff Manza.
\newblock Cleavage-based voting behavior in cross-national perspective:
  Evidence from six postwar democracies.
\newblock {\em Social Science Research}, 35(1):88--128, 2006.

\bibitem[\protect\citeauthoryear{Chen \bgroup \em et al.\egroup }{2013}]{f}
Jowei Chen, Jonathan Rodden, et~al.
\newblock Unintentional gerrymandering: Political geography and electoral bias
  in legislatures.
\newblock {\em Quarterly Journal of Political Science}, 8(3):239--269, 2013.

\bibitem[\protect\citeauthoryear{Dawkins}{2004}]{d}
Casey~J Dawkins.
\newblock Measuring the spatial pattern of residential segregation.
\newblock {\em Urban Studies}, 41(4):833--851, 2004.

\bibitem[\protect\citeauthoryear{Dawkins}{2007}]{e}
Casey~J Dawkins.
\newblock Space and the measurement of income segregation.
\newblock {\em Journal of Regional Science}, 47(2):255--272, 2007.

\bibitem[\protect\citeauthoryear{DeFord \bgroup \em et al.\egroup }{2020}]{g}
Daryl DeFord, Moon Duchin, and Justin Solomon.
\newblock A computational approach to measuring vote elasticity and
  competitiveness.
\newblock {\em Statistics and Public Policy}, (just-accepted):1--30, 2020.

\bibitem[\protect\citeauthoryear{Dey and Bhattacharyya}{2015}]{dey2015sample}
Palash Dey and Arnab Bhattacharyya.
\newblock Sample complexity for winner prediction in elections.
\newblock In {\em Proceedings of the 2015 International Conference on
  Autonomous Agents and Multiagent Systems}, pages 1421--1430, 2015.

\bibitem[\protect\citeauthoryear{Erd{\'e}lyi \bgroup \em et al.\egroup
  }{2015}]{3}
G{\'a}bor Erd{\'e}lyi, Edith Hemaspaandra, and Lane~A Hemaspaandra.
\newblock More natural models of electoral control by partition.
\newblock In {\em International Conference on Algorithmic DecisionTheory},
  pages 396--413. Springer, 2015.

\bibitem[\protect\citeauthoryear{Feldman \bgroup \em et al.\egroup
  }{2015}]{feldman2015certifying}
Michael Feldman, Sorelle~A Friedler, John Moeller, Carlos Scheidegger, and
  Suresh Venkatasubramanian.
\newblock Certifying and removing disparate impact.
\newblock In {\em proceedings of the 21th ACM SIGKDD international conference
  on knowledge discovery and data mining}, pages 259--268, 2015.

\bibitem[\protect\citeauthoryear{Geyik \bgroup \em et al.\egroup
  }{2019}]{yang2017measuring2}
Sahin~Cem Geyik, Stuart Ambler, and Krishnaram Kenthapadi.
\newblock Fairness-aware ranking in search \& recommendation systems with
  application to linkedin talent search.
\newblock In {\em Proceedings of the 25th acm sigkdd international conference
  on knowledge discovery \& data mining}, pages 2221--2231, 2019.

\bibitem[\protect\citeauthoryear{Grauwin \bgroup \em et al.\egroup
  }{2012}]{grauwin2012dynamic}
Sebastian Grauwin, Florence Goffette-Nagot, and Pablo Jensen.
\newblock Dynamic models of residential segregation: An analytical solution.
\newblock {\em Journal of Public Economics}, 96(1-2):124--141, 2012.

\bibitem[\protect\citeauthoryear{Guiver and Snelson}{2009}]{guiver2009bayesian}
John Guiver and Edward Snelson.
\newblock Bayesian inference for plackett-luce ranking models.
\newblock In {\em proceedings of the 26th annual international conference on
  machine learning}, pages 377--384, 2009.

\bibitem[\protect\citeauthoryear{Jagabathula and
  Shah}{2008}]{jagabathula2008inferring}
Srikanth Jagabathula and Devavrat Shah.
\newblock Inferring rankings under constrained sensing.
\newblock {\em Advances in Neural Information Processing Systems}, 21, 2008.

\bibitem[\protect\citeauthoryear{Jansen \bgroup \em et al.\egroup
  }{2013}]{jansen2013class}
Giedo Jansen, Geoffrey Evans, and Nan~Dirk De~Graaf.
\newblock Class voting and left--right party positions: A comparative study of
  15 western democracies, 1960--2005.
\newblock {\em Social science research}, 42(2):376--400, 2013.

\bibitem[\protect\citeauthoryear{Kuhlman and
  Rundensteiner}{2020}]{kuhlman2020rank}
Caitlin Kuhlman and Elke Rundensteiner.
\newblock Rank aggregation algorithms for fair consensus.
\newblock {\em Proceedings of the VLDB Endowment}, 13(12), 2020.

\bibitem[\protect\citeauthoryear{Kuhlman \bgroup \em et al.\egroup
  }{2021}]{kuhlman2021measuring}
Caitlin Kuhlman, Walter Gerych, and Elke Rundensteiner.
\newblock Measuring group advantage: A comparative study of fair ranking
  metrics.
\newblock In {\em Proceedings of the 2021 AAAI/ACM Conference on AI, Ethics,
  and Society}, pages 674--682, 2021.

\bibitem[\protect\citeauthoryear{Lasisi}{2018}]{6}
Ramoni~O Lasisi.
\newblock Improved manipulation algorithms for district-based elections.
\newblock In {\em The Thirty-First International Flairs Conference}, 2018.

\bibitem[\protect\citeauthoryear{Lev and Lewenberg}{2019}]{5}
Omer Lev and Yoad Lewenberg.
\newblock “reverse gerrymandering”: Manipulation in multi-group decision
  making.
\newblock In {\em Proceedings of the AAAI Conference on Artificial
  Intelligence}, volume~33, pages 2069--2076, 2019.

\bibitem[\protect\citeauthoryear{Lewenberg \bgroup \em et al.\egroup
  }{2017}]{1}
Yoad Lewenberg, Omer Lev, and Jeffrey~S Rosenschein.
\newblock Divide and conquer: Using geographic manipulation to win
  district-based elections.
\newblock In {\em Proceedings of the 16th Conference on Autonomous Agents and
  MultiAgent Systems}, pages 624--632, 2017.

\bibitem[\protect\citeauthoryear{Lin}{2010}]{lin2010rank}
Shili Lin.
\newblock Rank aggregation methods.
\newblock {\em Wiley Interdisciplinary Reviews: Computational Statistics},
  2(5):555--570, 2010.

\bibitem[\protect\citeauthoryear{Lu and Boutilier}{2011}]{lu2011learning}
Tyler Lu and Craig Boutilier.
\newblock Learning mallows models with pairwise preferences.
\newblock In {\em ICML}, 2011.

\bibitem[\protect\citeauthoryear{Lu and Boutilier}{2013}]{lu2013multi}
Tyler Lu and Craig Boutilier.
\newblock Multi-winner social choice with incomplete preferences.
\newblock In {\em Twenty-Third International Joint Conference on Artificial
  Intelligence}, 2013.

\bibitem[\protect\citeauthoryear{Mitra}{2020}]{mitra2020electoral}
Adway Mitra.
\newblock Electoral david vs goliath: How does the spatial concentration of
  electors affect district-based elections?
\newblock {\em arXiv preprint arXiv:2006.11865}, 2020.

\bibitem[\protect\citeauthoryear{Negahban \bgroup \em et al.\egroup
  }{2012}]{negahban2012iterative}
Sahand Negahban, Sewoong Oh, and Devavrat Shah.
\newblock Iterative ranking from pair-wise comparisons.
\newblock {\em Advances in neural information processing systems}, 25, 2012.

\bibitem[\protect\citeauthoryear{Pitman}{1995}]{crp}
Jim Pitman.
\newblock Exchangeable and partially exchangeable random partitions.
\newblock {\em Probability theory and related fields}, 102(2):145--158, 1995.

\bibitem[\protect\citeauthoryear{Qin \bgroup \em et al.\egroup
  }{2010}]{qin2010new}
Tao Qin, Xiubo Geng, and Tie-Yan Liu.
\newblock A new probabilistic model for rank aggregation.
\newblock {\em Advances in neural information processing systems}, 23, 2010.

\bibitem[\protect\citeauthoryear{Rajkumar and
  Agarwal}{2014}]{rajkumar2014statistical}
Arun Rajkumar and Shivani Agarwal.
\newblock A statistical convergence perspective of algorithms for rank
  aggregation from pairwise data.
\newblock In {\em International conference on machine learning}, pages
  118--126. PMLR, 2014.

\bibitem[\protect\citeauthoryear{Rajkumar \bgroup \em et al.\egroup
  }{2015}]{pairwise}
Arun Rajkumar, Suprovat Ghoshal, Lek-Heng Lim, and Shivani Agarwal.
\newblock Ranking from stochastic pairwise preferences: Recovering condorcet
  winners and tournament solution sets at the top.
\newblock In {\em International Conference on Machine Learning}, pages
  665--673. PMLR, 2015.

\bibitem[\protect\citeauthoryear{Relia}{2021}]{relia2021dire}
Kunal Relia.
\newblock Dire committee: Diversity and representation constraints in
  multiwinner elections.
\newblock {\em arXiv preprint arXiv:2107.07356}, 2021.

\bibitem[\protect\citeauthoryear{Schelling}{1971}]{schelling1971dynamic}
Thomas~C Schelling.
\newblock Dynamic models of segregation.
\newblock {\em Journal of mathematical sociology}, 1(2):143--186, 1971.

\bibitem[\protect\citeauthoryear{Stoica \bgroup \em et al.\egroup }{2019}]{4}
Ana-Andreea Stoica, Abhijnan Chakraborty, Palash Dey, and Krishna~P Gummadi.
\newblock Minimizing margin of victory for fair political and educational
  districting.
\newblock {\em arXiv preprint arXiv:1909.05583}, 2019.

\bibitem[\protect\citeauthoryear{Volkovs and Zemel}{2012}]{volkovs2012flexible}
Maksims~N Volkovs and Richard~S Zemel.
\newblock A flexible generative model for preference aggregation.
\newblock In {\em Proceedings of the 21st international conference on World
  Wide Web}, pages 479--488, 2012.

\bibitem[\protect\citeauthoryear{Xia}{2012}]{xia2012computing}
Lirong Xia.
\newblock Computing the margin of victory for various voting rules.
\newblock In {\em Proceedings of the 13th ACM conference on electronic
  commerce}, pages 982--999, 2012.

\bibitem[\protect\citeauthoryear{Yang and
  Stoyanovich}{2017}]{yang2017measuring}
Ke~Yang and Julia Stoyanovich.
\newblock Measuring fairness in ranked outputs.
\newblock In {\em Proceedings of the 29th international conference on
  scientific and statistical database management}, pages 1--6, 2017.

\bibitem[\protect\citeauthoryear{Zehlike \bgroup \em et al.\egroup
  }{2017}]{zehlike2017fa}
Meike Zehlike, Francesco Bonchi, Carlos Castillo, Sara Hajian, Mohamed Megahed,
  and Ricardo Baeza-Yates.
\newblock Fa* ir: A fair top-k ranking algorithm.
\newblock In {\em Proceedings of the 2017 ACM on Conference on Information and
  Knowledge Management}, pages 1569--1578, 2017.

\end{thebibliography}

\end{document}